\documentclass[sn-mathphys,Numbered,iicol,hyperref={colorlinks = true,linkcolor = blue}]{sn-jnl}

\usepackage{graphicx}%
\usepackage{multirow}%
\usepackage{amsmath,amssymb,amsfonts}%
\usepackage{amsthm}%
\usepackage{mathrsfs}%
\usepackage[title]{appendix}%
\usepackage{xcolor}%
\usepackage{textcomp}%
\usepackage{manyfoot}%
\usepackage{booktabs}%
\usepackage{algorithm}%
\usepackage{algorithmicx}%
\usepackage{algpseudocode}%
\usepackage{listings}%

\usepackage{graphics}
\usepackage{rotating}
\usepackage{ulem}
\usepackage{changes}
\usepackage{siunitx}
\usepackage{dcolumn}
\usepackage{bm}
\usepackage{rotating}
\usepackage{subcaption}

\theoremstyle{thmstyleone}%
\theoremstyle{thmstyletwo}%

\theoremstyle{thmstylethree}%

\newcommand{\s}{$S$}
\newcommand{\ssbar}{$S\bar{S}$}
\newcommand{\Suuddss}{$S$($uuddss$)}
\newcommand{\pbHethree}{$\bar{p}$\hspace*{0.4mm}--\hspace*{0.2mm}$^3$He}

\begin{document}

\let\pdfpageheight\paperheight
\let\pdfpagewidth\paperwidth

\title[pbar3He4DM]{Searching for a dark matter particle with anti-protonic atoms}

\pagestyle{plain}
\setcounter{page}{1} 
 
\author*[1]{\fnm{Michael} \sur{Doser}}\email{michael.doser@cern.ch}

\author[2]{\fnm{Glennys} \sur{Farrar}}\email{glennys.farrar@nyu.edu}
\equalcont{These authors contributed equally to this work.}

\author[3]{\fnm{Georgy} \sur{Kornakov}}\email{georgy.kornakov@cern.ch}
\equalcont{These authors contributed equally to this work.}

\affil*[1]{\orgname{CERN}, \orgaddress{\street{Esplanade des Particules 1}, \city{Geneva}, \postcode{1211}, \country{Switzerland}}}

\affil[2]{\orgdiv{Center for Cosmology and Particle Physics}, \orgname{New York University}, \orgaddress{ \city{New York}, \postcode{10003}, \state{NY}, \country{USA}}}

\affil[3]{\orgname{Warsaw University of Technology}, \orgaddress{\street{ul. Koszykowa 75}, \city{Warsaw}, \postcode{00-662}, \country{Poland}}}


\abstract{A wide range of dark matter candidates have been proposed and are actively being searched for in a large number of experiments, both at high (TeV) and low (sub meV) energies.  
One dark matter candidate, a deeply bound $uuddss$ sexaquark, \s, with mass $\sim 2$ GeV (having the same quark content as the hypothesized H-dibaryon, but long lived) is particularly difficult to explore experimentally.
In this paper, we propose a scheme in which such 
a state could be produced at rest 
through the formation of \pbHethree\ antiprotonic atoms and their annihilation into \s\ + $K^+K^+\pi^-$,
identified both through the unique tag of a S=+2, Q=+1 final state, as well as through full kinematic reconstruction of the final state recoiling against it.}




\maketitle


\section{Introduction}

The identification of the nature of dark matter is one of the most pressing concerns of particle physics. Candidates for dark matter particles cover an exceedingly wide range, with masses predicted to lie between 10$^{-22}$ eV and 10$^{10}$ eV, depending on the specific type of candidate and not including primordial black holes. Searches for candidates with masses above the GeV scale are generally accelerator-based: recent reviews of these can be found in~\cite{accelerator_searches_1, accelerator_searches_2}. Putative low-mass dark matter candidates (e.g. axions, ALP's, chameleons, ...), whose detection requires highly sensitive devices, among them quantum sensors which can detect discrete state changes from one quantum state to another that interactions with dark matter can trigger, are equally actively being searched for~\cite{ultralightDM, axionDM}. One 
dark matter candidate, the sexaquark\textcolor{black}{~(S)}~\cite{stableS,fudsDM18}, would however not have been detected by either approach, while prior searches for similar states in the GeV region  (notably, the H-dibaryon~\cite{Jaffe}) would not have been sensitive to the \s~\cite{fS22}. 

The sexaquark is a hypothesized deeply bound, long-lived or stable $0^+$ state of $uuddss$ quarks with B=+2, S=--2 and Q=0.  The quark content and quantum numbers are the same as the H-dibaryon, however the properties are distinctly different since the H-dibaryon was envisaged to be relatively weakly bound with a lifetime $\mathcal{O}(10^{-10}$s).  Accelerator experiments have excluded the existence of the H-dibaryon, unless it is a weakly bound ``molecule" like the deuteron, but an \s\ would not have been discovered by experiments to date (see~\cite{fS22} for a survey and suggestions for strategies).  Three attributes of a stable \s\ make it very difficult to detect:\\ 
\begin{itemize}
\item The \s\ is neutral and a flavor singlet, so it does not couple to photons, pions and most other mesons, nor does it leave a track in a detector. \\
\item The \s\ has no pion cloud and is expected to be more compact than ordinary baryons.  This means the amplitude for interconversion between \s\ and baryons is small. \\
\item The mass of the \s\ makes it difficult to distinguish from the much more copious neutron. 
\end{itemize}

If $m_S < m_\Lambda + m_p + m_e = 2.054 $ GeV, its decay must be doubly-weak and its lifetime would be greater than the age of the Universe~\cite{fzNuc03}, making it an interesting dark matter candidate.  Direct detection experiments have not yet probed the relevant mass range with sufficient sensitivity to have excluded this possibility~\cite{fS22}. It has also been shown that the stability of neutron stars is not affected by the existence of sexaquarks, due to quark deconfinement~\cite{Shahrbaf+22}.
In addition to its significance as a Standard Model dark matter candidate, the existence of the \s\ and production of \ssbar\ pairs in $e^+ e^-$ collisions could make an undetected contribution to $R_{\rm had}$ and increase the Hadronic Vacuum Polarization (HVP)~\cite{f_g-2_22}. 
\textcolor{black}{
This would reduce at some level the tension between the dispersive determination of the HVP based on experimental measurements of $R_{\rm had}$~\cite{Aoyama+20,colangelo+22} and the predictions of lattice QCD~\cite{hadronvacuumpolarization,blumUKQCD23}, which are much closer to the value needed to explain the measured muon anomalous magnetic moment~\cite{g-2_prl23} (Note however that the recent CMD-3 data has greatly reduced the discrepancy between the dispersive value and g-2~\cite{CMD-3sept23} and lattice QCD predictions~\cite{blumUKQCD23};  moreover the change in g-2 from \ssbar\ production in $e^+ e^-$ collisions is expected to be small).} These considerations, combined with the importance of fully exploring the hadron spectrum, makes the existence of the \s\ an important question even if the \s\ does not explain Dark Matter.

Formation of multiquark states (e.g.~tetraquarks or pentaquarks~\cite{pentaquarks}) is well established in high energy particle interactions~\cite{multiquark}, with their dense environment of numerous quark-antiquark pairs, and such a production channel for sexaquarks has also been proposed and investigated~\cite{Farrar_LHC1, Farrar_LHC2}. While in such high energy collisions, production rates of multiquark states may well be high, detection of all involved participants and complete kinematical reconstruction of the potentially produced sexaquark is very difficult, given the generally high multiplicity of participants and the low mass of the proposed sexaquark. In contrast, production at low energy, which facilitates full reconstruction, requires establishing a situation with high numbers and high local density of quarks, and in which furthermore, at least two strange quarks are produced. While searches in e.g.~$\Upsilon$ decays have been attempted~\cite{PhysRevLett.110.222002,PhysRevLett.122.072002}, annihilation of antiprotons (or antineutrons) with nuclei, with a concomitant larger number of quarks and antiquarks, may have a better potential to achieve these conditions. On one hand, annihilation between an antiproton and a nucleon results - in the naive quark model - in the interaction of three quark-antiquark pairs, leading potentially to the formation of multiple strange-antistrange pairs, as has been pointed out in proposals to investigate double-strangeness production with antiprotons~\cite{double-strangeness}. On the other hand, the spectator nucleons potentially provide additional light quarks; such multi-nucleon interactions involving antiprotons have been observed~\cite{Riedlberger} 
in antiproton-nucleus annihilations at rest and have been studied theoretically~\cite{Cugnon}. 
This strategy has been applied in the past to (unsuccessful) searches for similar states, in particular for the H dibaryon~\cite{Jaffe} with identical quark composition as the \Suuddss, but with different mass, physical structure, and thus decay characteristics. In particular, previous searches for the H-dibaryon (with its $\Lambda\Lambda$ quark composition) have either relied on decays into $\Lambda\Lambda$, $\Lambda p \pi$ or $\Lambda n$  
~\cite{ALICE_femtoscopy, Takahashi, ALICE_2016267} or on exclusive co-production with $\bar{\Lambda}\bar{\Lambda}$ in $\Upsilon$ decays~\cite{PhysRevLett.110.222002} (requiring formation and binding of exactly 12 quarks and antiquarks simultaneously, with an estimated branching fraction of $\mathcal{O}(10^{-11})$~\cite{fS22}).
Here, we propose a novel approach that, while having similarilarities to the above, would lead to a very clean experimental tag for the production of stable sexaquarks in a low energy environment characterized by a high local density of quarks and antiquarks. We will investigate in the following a specific antiproton-nucleus system, first conceptually, to extract the salient characteristics of a reaction potentially resulting in production of a $uuddss$ sexaquark, and then investigate with a simulation the sensitivity of such a production process against potential background processes. 

\section{Formation process}
Our proposed scheme for producing $S(uuddss)$ sexaquark states at low energy starts
with antiprotonic atoms of $^3$He. Simply considering the masses of the involved participants results in a total energy budget of $m_{\bar{p}} + m_p + m_p + m_n$ $\sim$ 3750 MeV for a system at rest.  
As we shall show below, the formation of \Suuddss\ in the process \pbHethree~$\rightarrow S  K^+ K^+ \pi^- $ should be detectable for relative production rates down to $10^{-9}$ of all annihilations, with a clean experimental signature that allows determining precisely the mass of the (invisible) recoiling dark matter candidate \Suuddss\  system.

Annihilation of the $\bar{p}$ with an $n$ or a $p$ of the nucleus results - in the naive quark model - in quark annihilation and/or quark rearrangement; for simplicity, we will first focus on the case of $\bar{p}p$ annihilation with no additional spectator nucleons. If two quark-antiquark annihilations occur, then two $s\bar{s}$ pairs can be produced (albeit with low probability, and ignoring energy limitations in a first step), with either the $u\bar{u}$ or the $d\bar{d}$ surviving, in other words, $uud$ + $\bar{u}\bar{u}\bar{d}$ $\rightarrow$ $uss\bar{u}\bar{s}\bar{s}$  or  $dss\bar{d}\bar{s}\bar{s}$. Such processes, leading to the production of $\phi\phi$ pairs, have been observed to occur in $\bar{p}p$ annihilations in flight~\cite{BERTOLOTTO1995325}, once the available rest-frame energy is sufficient.

\begin{figure}
    \centering
    \includegraphics[width=0.44\textwidth]{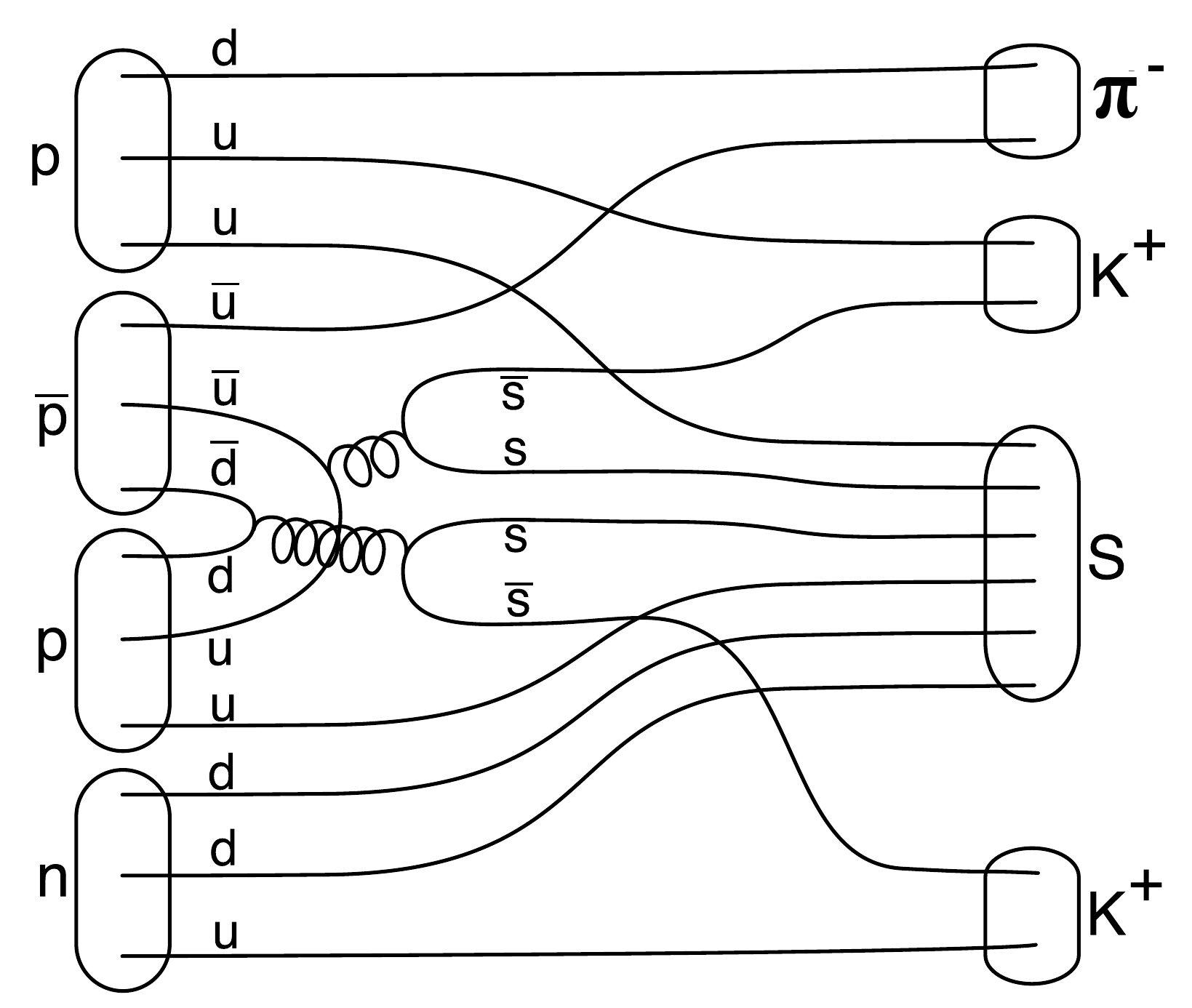}
    \vspace*{3mm}
    \caption{Quark rearrangement and annihilation graph for the formation of a $uuddss$ sexaquark state in \pbHethree~annihilations. S denotes the putative \Suuddss\ sexaquark state.}
    \label{fig:formation}
\end{figure}

Since even the lightest final states containing $\bar{s}\bar{s}ss$ quark combinations (4$K$, $\Xi^0\bar{\Xi^0}$, $\phi\phi$ and $\phi K K$) all exceed 2m$_p$, energy conservation requires that for annihilation at rest, this process occur together with at least one additional nucleon, either a $p$ or an $n$. For \pbHethree, this thus requires a three-nucleon interaction with a single spectator nucleon (in the case of the $\bar{p}d$ subsystem of \pbHethree) or a four-nucleon interaction (in the case of \pbHethree); here, we will consider the later process in the form of the quark picture involving all four nucleons (Fig.\ref{fig:formation}).

 For multi-nucleon annihilations, the question of whether cross-nucleon quark clusters can form is of relevance. In a study of deep-inelastic electron scattering from $^3$He~\cite{PhysRevLett.46.1376}, a geometric probability for a six-quark cluster configuration to occur in the $^3$He nucleus was found to be 16\% (and even that of a 9-quark cluster remains non negligible, at 1\%). 
In deuterons, the same probability is estimated to be 10\%~\cite{veselov1985realistic}; theoretical considerations~\cite{Cugnon} indicate both the production rate of individual two-body final states at a level of O($10^{-5}$) as well as an enhanced strange particle yield.

Not surprisingly, corresponding experimental indications for multi-nucleon annihilations have been found in several systems, such as the process $\bar{p}d \rightarrow p \pi$ which requires all nine (anti)quarks of the system to interact, and has a branching ratio of $10^{-5}$ \cite{Riedlberger}. Similarly, three-nucleon processes with additional spectator nucleons have been argued on the basis of the high yield of $\Lambda$ production, both in deuterium and in nitrogen\cite{Riedlberger, Bizzari}. Similarly, while searching for deeply bound $\bar{K}\bar{K}$-nuclear  states in $\bar{p}^4$He annihilations, the OBELIX experiment has identified final states containing two $K^+$, accompanied by two $\pi^-$ and missing \textcolor{black}{momentum and energy}, implying necessarily the involvement in the annihilation of more than one nucleon. 
This final state was interpreted as $2K^+\Sigma^-\Sigma^- p_s \rightarrow 2K^+2\pi^-(2n~p_s)$ with a yield of $(0.172 \pm 0.038)$ $\times$ 10$^{-4}$ \cite{Obelix_KpKp}.
Annihilation processes involving multiple nucleons are thus well established, with production rates in the order of $10^{-6} \sim 10^{-4}$ and with indications of the presence of multi-nucleon annihilations.

In the case of \pbHethree~(Fig.\ref{fig:formation}), after annihilation of two of the antiquarks of the $\bar{p}$ with two of the quarks of other nucleons to produce two $\bar{s}s$ pairs, the resulting intermediate quark composition  is then
$[u s s \bar{u} \bar{s} \bar{s}] [ d d u] [ u u d]$  or  $[d s s \bar{d} \bar{s} \bar{s}] [ d d u] [ u u d]$, depending on which quarks of the antiproton-proton pair annihilate (the former occurring two times more frequently than the later).
Pairing off the 3$^{rd}$ and 4$^{th}$ nucleons’ quarks results in
$[ u s s    d d u ] [ \bar{u} d ] [\bar{s} u ] [ \bar{s} u ]$  (in 2/3 $\times$ 1/3 of the cases), in $[ d s s    u u d ] [ \bar{d} u ] [ \bar{s} d ] [ \bar{s} d ]$ (in 1/3 $\times$ 1/3 of the cases), in
$[ u s s    d d u ] [ \bar{u} u ] [\bar{s} d ] [ \bar{s} u ]$ (in 2/3 $\times$ 2/3 of the cases) and in  $[ d s s    u u d ] [ \bar{d} d ] [ \bar{s} d ] [ \bar{s} u ]$ (in 2/3 $\times$ 2/3 of the cases).

The outcome is then production of the (putative) \Suuddss\ state in coincidence with two kaons and one pion:

\vspace*{-5mm}
\begin{align}
\textrm{\it{(uuddss)}} +  \pi^- + K^+ + K^+ \textrm{  (in } 2/3 \times  1/3 \textrm{ of the cases)} \\
\textrm{\it{(uuddss)}} +  \pi^+ + K^0 + K^0 \textrm{  (in } 1/3 \times  1/3 \textrm{ of the cases)} \\
\textrm{\it{(uuddss)}} +  \pi^0 + K^0 + K^+ \textrm{  (in } 2/3 \times 2/3 \textrm{  of the cases)} \\
\textrm{\it{(uuddss)}} +  \pi^0 + K^0 + K^+ \textrm{  (in } 1/3 \times 2/3 \textrm{ of the cases)} 
\end{align}

\begin{figure}
    \centering
    \includegraphics[width=0.4\textwidth]{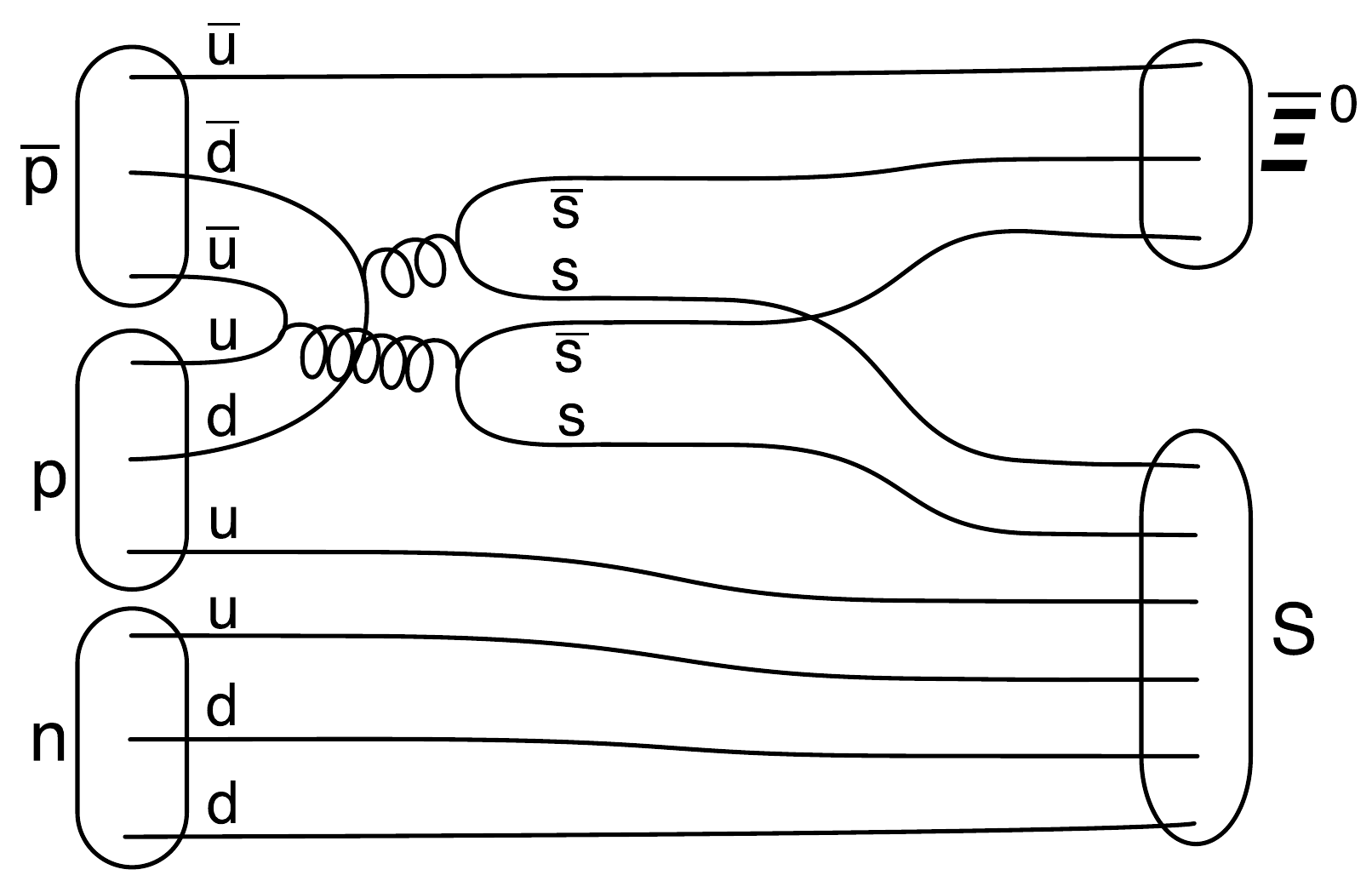}
    \hspace*{5mm}
    \includegraphics[width=0.4\textwidth]{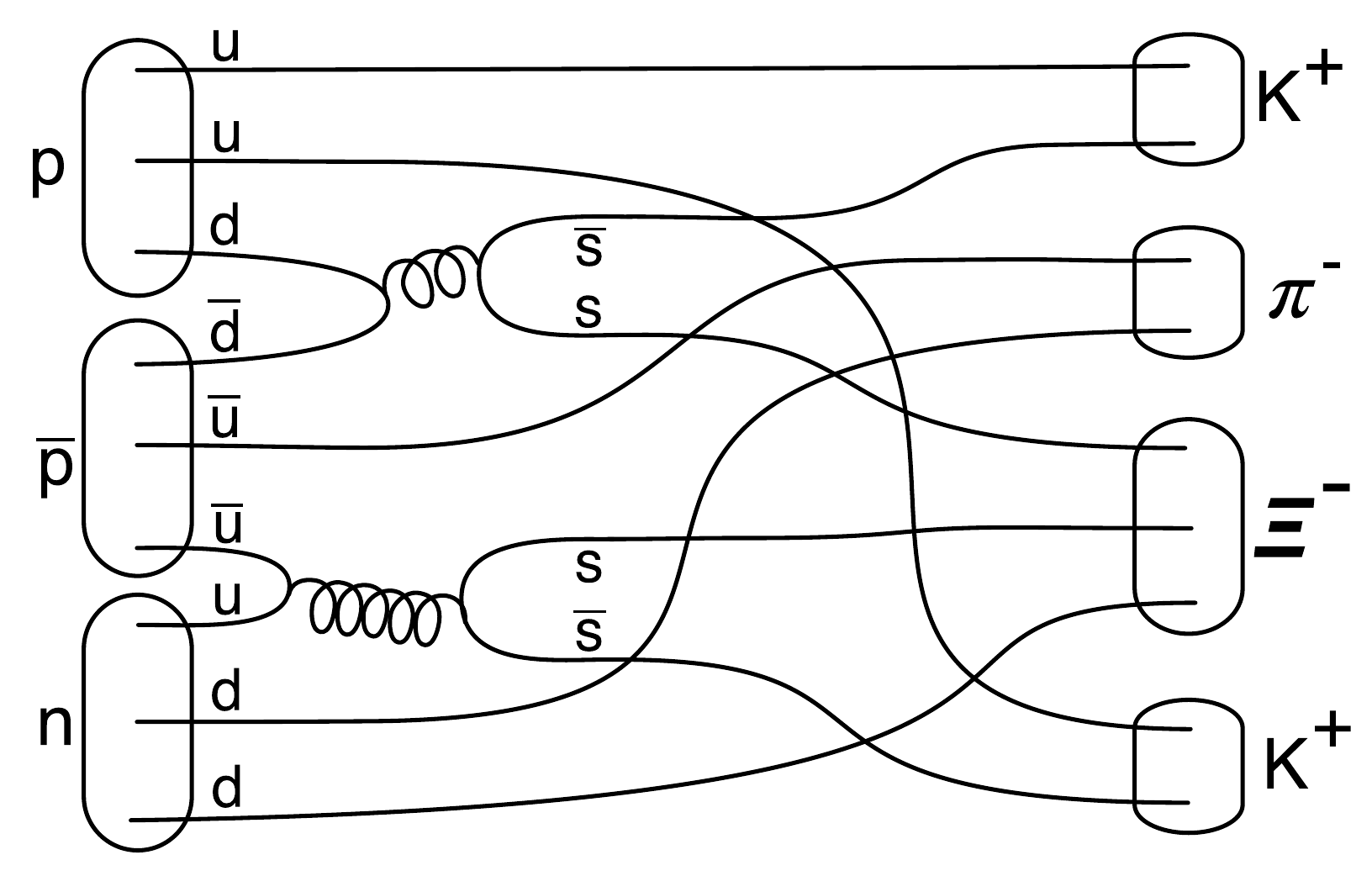}
   \vspace*{3mm}
    \caption{Three-nucleon \textcolor{black}{rearrangement and annihilation} graphs for $\bar{p}$d leading to the formation of a final state containing the $K^+K^+\pi^-$ topology or of the \Suuddss. In both cases, the spectator p would also form part of the final state. 
    Top: graph for $\bar{p}d \rightarrow \bar{\Xi^0} S$.
    Bottom: $\bar{p}d \rightarrow \Xi^- K^+K^+\pi^-$.
    In \pbHethree~annihilations, this channel, in which the $\Xi^-$ can be fully reconstructed via its $\Xi^- \rightarrow \Lambda \pi^-, \Lambda \rightarrow p \pi^-$ decay, and which is accompanied by a spectator proton $p_s$, is a useful cross check for the presence of multinucleon annihilation, but also for triggering, tracking and particle identification capabilities of the detector used to search for \Suuddss. }
    \label{fig:formation_2}
\end{figure}

In all cases, the total strangeness of the final state mesons is +2, while the sum of charges
$\Sigma_Q$ = +1.  
Final state masses sum to m(S) + 2m(K) + m($\pi$), allowing producing $S$ with $m_S$ $\in [0, 2600]$ MeV. The experimentally cleanest tag is (1), since all final state particles (except for the invisible \Suuddss\ state) are easily detectable and uniquely identifiable. Furthermore, kinematic reconstruction of the missing mass is unambiguously possible, since the initial energy and momentum of the \pbHethree~system are known. Note that also $\bar{p}d$ annihilation  can lead to the production of the sexaquark ($\bar{p}d \rightarrow \bar{\Xi}^0 S$, Fig. \ref{fig:formation_2}, left), but results in a much reduced mass window (m$_S<$ 1500 GeV) than is the case for \pbHethree~and is experimentally much more difficult to detect due to the dominant $\bar{\Xi}^0 \rightarrow \bar{\Lambda}\pi^0, \bar{\Lambda} \rightarrow \bar{p}\pi^+$ decay. 


We now discuss a number of standard model processes that result either in similar final state topologies, or that produce final states containing the characteristic $K^+K^+\pi^-$ topology (with additional mesons or baryons, but which might escape detection in a real detector) to determine whether these could result in potential backgrounds. No standard model processes result in the signal topology with a single missing particle; in all cases, particle misidentification plays a role in potential backgrounds:
\begin{itemize}

    \item The most challenging background channel stems from $\bar{p}p(d) \rightarrow K^+\bar{K^0}\pi^+\pi^-(d)$, with the $\pi^+$ being misidentified as a $K^+$ and both the $\bar{K^0}$ and the spectator $d$ escaping detection. This channel with a production rate of O(10$^{-3}$) will be studied in more detail below (section 3).

    \item Additional potential backgrounds stem from incomplete detector solid angle coverage. Annihilation of antiprotons on a deuteron-like two-nucleon system inside $^3$He with the fourth nucleon (a proton $p_s$) remaining as spectator, 
    $\bar{p}d(p_s) \rightarrow \Xi^- K^+ K^+ \pi^- (p_s)$ is kinematically allowed (Fig.~\ref{fig:formation_2}, right). Here, the decay of the $\Xi^- \rightarrow \Lambda \pi^- ; \Lambda \rightarrow n\pi^0$ produces a recoiling $\pi^-$, a neutron and two photons. Misidentification of this topology as signal requires missing the $\pi^-$ of the $\Xi^-$ decay and the spectator proton, but also the missing mass recoiling against the $K^+ K^+ \pi^-$ system will not result in a unique recoil mass.

 \item A related background with a spectator neutron ($n_s$) stems from:
  $\bar{p}d(n_s) \rightarrow \Lambda^0 K^+ K^+ K^- (n_s)$, which is is kinematically allowed, as is the related process $\bar{p}d(n) \rightarrow \Sigma^0 K^+ K^+ K^- n$. While mis-identification of the $K^-$ as $\pi^-$ can lead to the signal topology, the missing mass recoiling against the $K^+ K^+$``$\pi^-$'' system will not result in a unique recoil mass.

  \item A more challenging, also related, background stems from  
    $\bar{p}pp(n_s) \rightarrow \Xi^0 K^+ K^+ \pi^- (n_s)$. Fully neutral decays of the $\Xi^0 \rightarrow \Lambda \pi^0 ; \Lambda \rightarrow n\pi^0$ produce a recoiling neutron and four photons, in addition to the spectator neutron. Here, only the missing mass recoiling against the $K^+ K^+ \pi^-$ system, which will not result in a unique recoil mass, can be used to discriminate this potential background.

\item $\bar{p}n + p + p$, with $\bar{p}n \rightarrow K^-K^+\pi^-$. This background requires both misidentification of the $K^-$ as $K^+$ as well as both protons escaping detection.




\end{itemize}

To summarize: the potential backgrounds to the signal, either via rescattering or multi-nucleon annihilation processes, cannot result in a $K^+ K^+ \pi^-$ final state recoiling against a single state;
only particle misidentification (or particles escaping detection) can result in such a final state. 
In the later case, the resulting topology requires a study of the missing mass distribution to establish down to which level sensitivity to the searched for \Suuddss\ is feasible. The fake and misidentification rates are evaluated in the following with a simple Monte Carlo simulation and using known branching ratios for $\bar{p}p$ and $\bar{p}n$ annihilations into final states with at least one K$^+$.



\section{Simulations}
\begin{figure}
    \centering
    \includegraphics[width=0.4\textwidth]{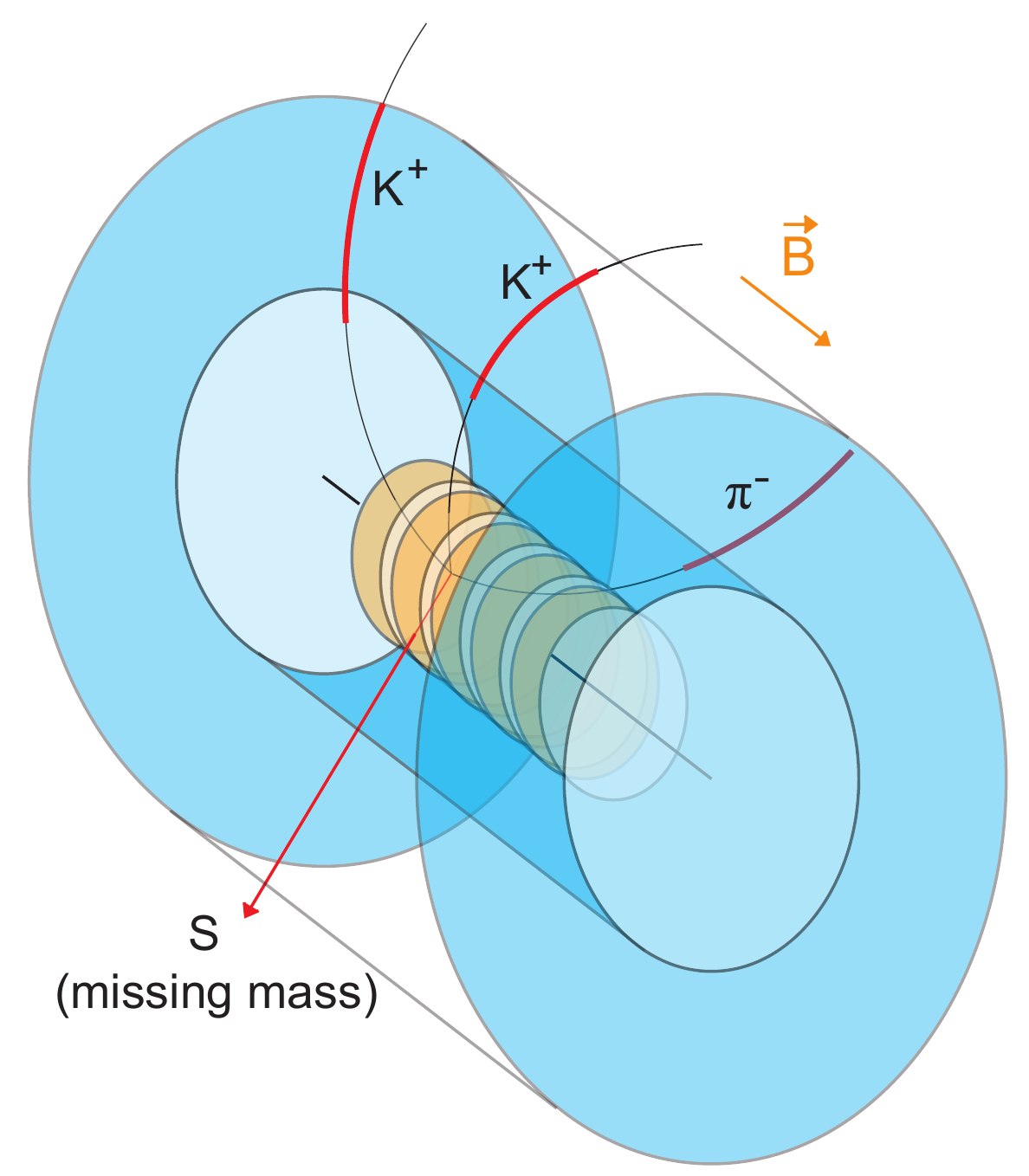}
   \vspace*{3mm}
    \caption{\textcolor{black}{Sketch of the searched-for $\pi^-K^+K^+ mm$ topology (where $mm$ is the recoiling missing mass) with a generic compact cylindrical TPC-like tracking detector surrounding Penning trap structures within which the formation and annihilation process would occur. The overall device would be embedded in a multi-T solenoidal magnetic field.}}
    \label{fig:schematic_detector}
\end{figure}

The \pbHethree~annihilation process has been studied using the Geant4 simulation framework~\cite{GEANT4:2002zbu,Allison:2006ve,Allison:2016lfl},
obtaining an estimate of the production rate of the topologies of interest with final states containing $K^+ K^+ \pi^-$ and possible backgrounds to the process.
In addition, for rare multi-nucleon processes we have studied the final states using Monte-Carlo phase space generators implemented in the ROOT framework~\cite{BRUN199781,James:275743}. \textcolor{black}{Figure~\ref{fig:schematic_detector} provides a sketch of the searched-for topology within a generic tracking and particle identification detector, such as a compact TPC.}

The annihilation at rest of antiprotons on nuclei is simulated in several steps, following the procedure as in~\cite{antiprotonic_atoms}. 
First, 1~keV antiprotons are shot at a nm-thin $^{3}$He target where via electromagnetic interaction they are slowed down until captured by an atom. 
The thin target prevents new interactions of any produced states with the bulk material, providing similar conditions to those of ultra-high vacuum. Then, the annihilation of the antiproton with one of the nucleons is described by the Geant4 FTFP\_BERT\_HP physics list~\cite{antimatter_geant4}, which assumes a quark gluon string model for high energy interactions. 

After each capture and annihilation event, information on the produced final state particles such as their type, four-momentum vector, and charge is accessible for subsequent analyses. Final states containing $K^+ K^+ \pi^-$ candidates are searched for in view of subsequent analysis. 
Zero events with this pure final state were found in 10$^9$ simulated annihilations. 
This is expected due to the lack of rare multi-nucleon annihilation processes in the code. 
However, similar final states are possible when there is a wrong tagging of the particle type (misidentification) or some of the final state particles are not measured (inefficiency) in the experiment. 

For simulating these effects we have assumed a minimal momenta for particles for being reconstructed: 10~MeV/$c$ for $\pi$, 25~MeV/$c$ for K, 40~MeV/$c$ for protons, and 80~MeV/$c$ for deuterons. The inefficiency of the tracking device is of 0.5 \%, momentum resolution ($\Delta p/p$) is set to 1 \%, and a probability of wrong particle type assignment (confusing a pion with a kaon and vice-versa) is set to 0.5 \%. With those assumptions, we have analysed the final states containing reconstructed $K^+ K^+ \pi^-$ triplets. 

The analysis of the tagged final states produced in Geant4 shows that most of the background events have the following origin sorted from most likely to less likely:
\begin{enumerate}
    \item lost low energy deuteron or proton with one negative pion and two positive pions misidentified as kaons. 
    \item lost low energy proton or deuteron, kaon and two pions, one misidentified as the second kaon.
    \item inefficient reconstruction of proton/deuteron, kaon and two pions, one of them misidentified.
\end{enumerate}

\begin{figure}[hbtp]
    \centering
    \includegraphics[width=0.51\textwidth]{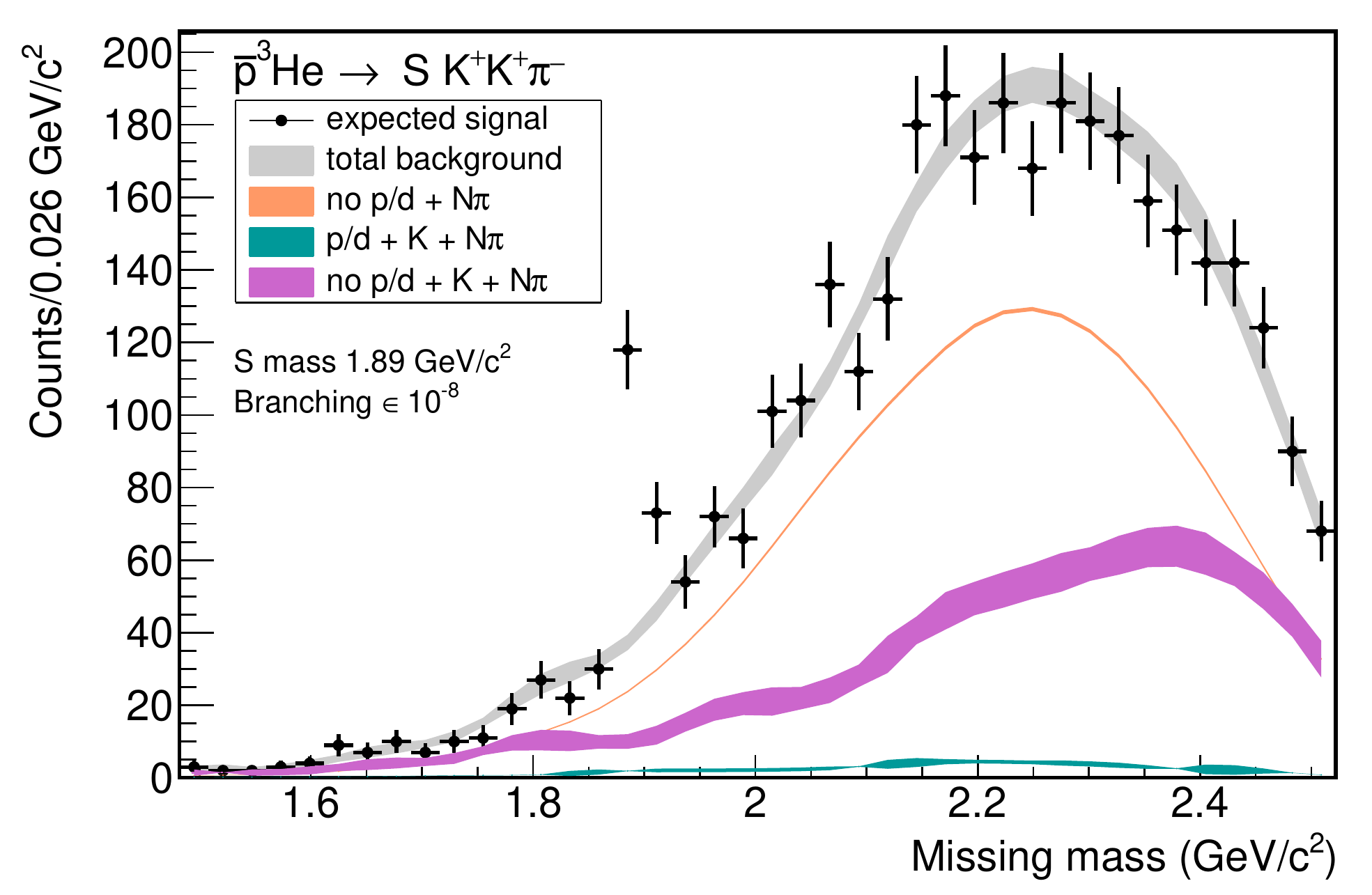}
    \includegraphics[width=0.51\textwidth]{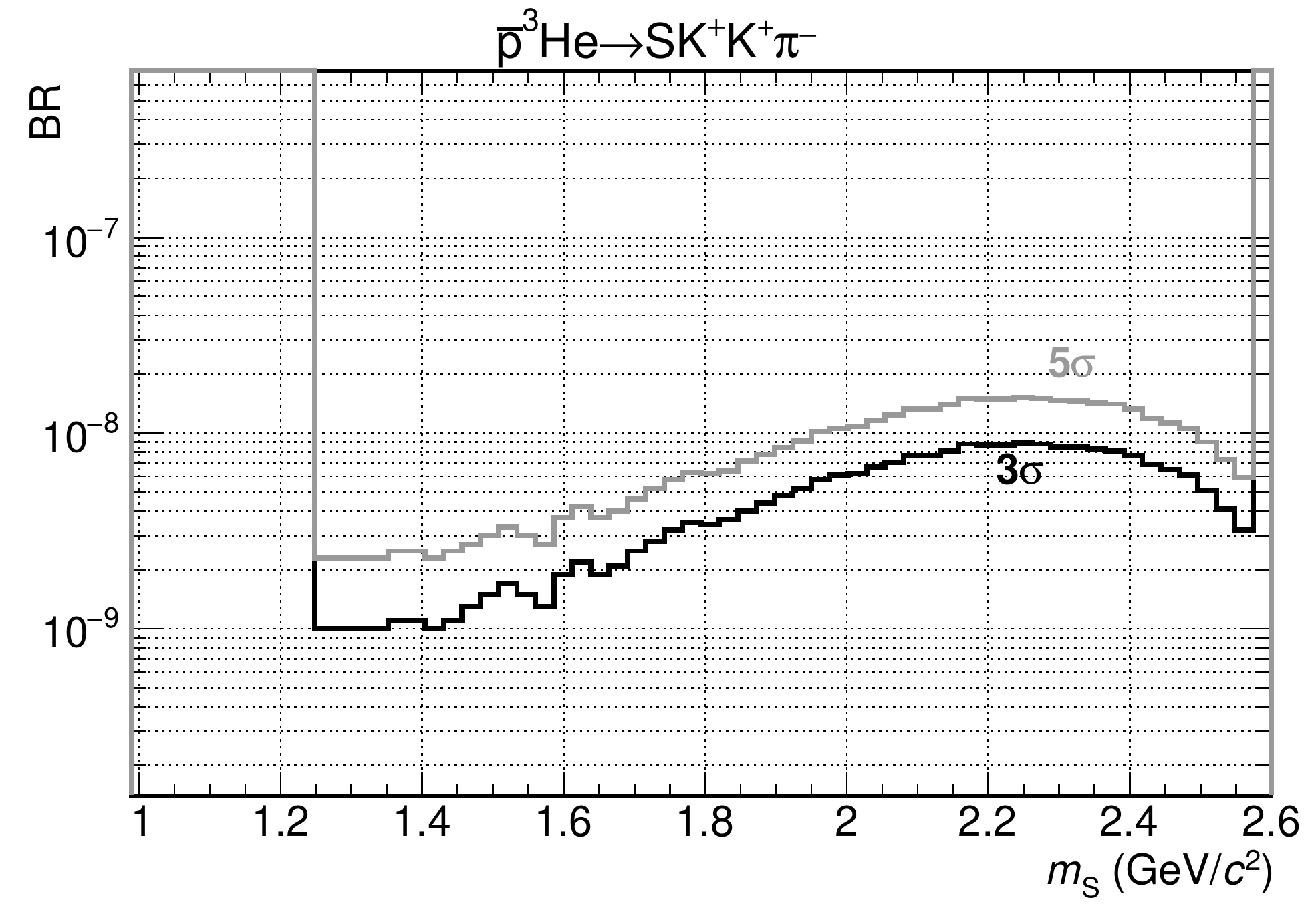}
    \caption{Top: Reconstructed missing mass distribution from 10$^{10}$ simulated events in Geant4 of the $K^+K^+\pi^-$ system in \pbHethree~$\rightarrow uuddss + K^+ + K^+ + \pi^-$ together with background. The peak at 1.89 GeV/c$^2$ corresponds to production of \Suuddss\ with a branching ratio of 10$^{-8}$. 
    Bottom: Experimental sensitivity (3 and 5-$\sigma$) to the branching ratios of \Suuddss\ in  \pbHethree~$\rightarrow uuddss + K^+ + K^+ + \pi^-$ as a function of $m_S$ for 10$^{10}$ events.}
    \label{fig:mm2}
\end{figure}

The expected background distributions for 10$^{10}$ antiproton annihilations on $^3$He and the above main sources are shown in Fig.~\ref{fig:mm2}. The condition of full kinematic reconstruction of the final state is able to reject most of the events with similar topologies. In the mass region around the deuteron, we expect to have 60--70 background counts per 0.052 GeV/c$^2$, increasing towards higher masses, reaching approximately 350 counts at the maximum at 2.2 GeV/c$^2$. In the same figure the expected \s~signal distribution after smearing the momentum of the reconstructed tracks is also included for an assumed production rate in \pbHethree\ annihilations of 10$^{-8}$ and a mass of 1.89 GeV/c$^2$. The observed width for a stable \s~is only due to reconstruction and tracking uncertainties and is expected to be in the order of 0.01~GeV/c$^2$. 

In addition to the Geant4 simulations we have performed checks for the rare multi-nucleon processes with branching ratios below 10$^{-5}$~\cite{double-strangeness}. 
In order to study the expected background distribution in events with two kaons produced together with $\Lambda$ or $\Xi$ baryons, we studied the missing mass distribution assuming an inefficient reconstruction of the weakly decaying particles. 
The missing mass distribution starts at 2.2 GeV/c$^2$, thus not representing a real source of background for the \s~searches. 

We also used the phase-space generator to test the missing mass distribution for a possible channel including in addition to the \s, two kaons and a charged pion further neutral pions. In this case, the sharp peak in the missing mass disappears and it is shifted towards higher values. Thus, this potential source of background will not contribute to the expected signal region. 

The background distribution obtained using the Geant4 framework has been used as the basis to estimate the sensitivity to different branching ratios of \s~in an experiment. An example of the \s~with a mass of 1890 MeV/c$^2$ and a branching ratio of 10$^{-8}$ is shown in Fig.~\ref{fig:mm2}. 

The simulated points then are used to estimate the sensitivity using the Asimov significance test~\cite{Cowan:2010js} for 10$^{10}$ annihilation events. The resulting 3 and 5-$\sigma$ sensitivities to branching ratios as a function of the \s~mass ($m_{S}$) are shown in the right panel of Fig.~\ref{fig:mm2}. The highest sensitivity, to a branching ratio of $10^{-9}$ is in the interval between 1200 and 1600 MeV/c$^2$, decreasing to $10^{-8}$ for masses around 2200 MeV/c$^2$. In the most favourable mass region between 1800 and 2000 MeV/c$^2$ the predicted sensitivity is 3-4 $\times 10^{-9}$. 

The simulated number of annihilations is equivalent to an experiment lasting for a year at the Antiproton Decelerator (AD) at CERN providing 100 keV antiprotons in bunches of 10$^{6\sim 7}$ particles every 100~s~\cite{Autin:1995iv,ELENA:2018hyb}. In order to ensure the detection of any of the recoiling deuterons (or protons) which are the main source of background events the experiment should use a light target, preventing absorption of the final-state particles. For this reason, reconstruction using liquid or gaseous targets would be very difficult. 

An alternative is provided by the formation of \pbHethree~in Penning traps~\cite{antiprotonic_atoms}. 
In-trap formation in ultra high vacuum (UHV) ensures that any low transverse energy nuclear remnants can be detected and identified axially (e.g.~via a micro channel plate (MCP) and time-of-flight detectors), while higher energy mesons can be detected via a traditional cylindrical tracking detector embedded within the Penning trap's solenoidal magnetic field which allows to reconstruct accurately the momentum of any produced charged particles \textcolor{black}{(Figure~\ref{fig:schematic_detector})}.

\section{Discussion} \label{Discussion}

The above discussion demonstrates that \Suuddss\ sexaquarks, a specific dark matter candidate, can potentially be formed and be detected with high sensitivity and over a very low background in antiproton-nuclear annihilations at rest, with the most appropriate system for detection being \pbHethree~annihilations. This process assumes that re-arrangements of quarks can happen across nucleon boundaries. Experimentally, such multi-nucleon interactions have been shown to occasionally occur in antiproton annihilation~\cite{Riedlberger, Montagna}. 
The same process could also occur in heavier nuclei than $^3$He,
resulting in final states of the type $(uuddss) + K^+K^+\pi^- + n \times [uud] + m \times [udd]$. 
Kinematically, 
such a state with supernumerary protons or neutrons might decay into $\Lambda\Lambda N’$ 
or would result in multiple particles recoiling 
against the $K^+K^+\pi^-$ system, 
and thus entail a less clean final state topology than for \pbHethree. 
Finally, while the  formation of the sexaquark system in antiproton--tritium interactions should proceed as in $^3$He, the resulting final state does not have the same characteristic S=+2 signature.

Formation of the \pbHethree~antiprotonic atom through co-trapped ionic precursor systems, e.g. $\bar{p}$ + $^3$He$^+$, can rely either on pulsed formation via charge exchange between co-trapped antiprotons and anionic atomic ions that are first photo-ionized and then immediately excited into Rydberg states~\cite{Doser_review}, similar to the pulsed formation of antihydrogen~\cite{Aegis}, requiring production of metastable $^3$He$^-$~\cite{PhysRevA.47.890}, or on three-body interactions between antiprotons and a trapped ensemble of cationic $^3$He$^+$ atomic ions, similarly to the manner in which antihydrogen atoms can be formed~\cite{Athena, Atrap}. 

\textcolor{black}{Alternatively, replacing $^3$He with tritium would result in an $(S \pi^- K^+ K^0$) final state in a process similar to that of Figure 1. Contrary to antiproton annihilation on $^3$He to $(S \pi^- K^+ K^+$) however, which has no standard model background with the same final state topology, for antiproton annihilation on tritium the standard model process $\bar{p}p(n n) \rightarrow \pi^- K^+ \bar{K^0} (n n)$ with two spectator neutrons occurs at a branching ratio level of O(10$^{-3}$).}

In summary, we have shown that formation of \Suuddss\ in the process \pbHethree~$\rightarrow S  K^+ K^+ \pi^- $ should be detectable for relative production rates down to $10^{-9}$ of all annihilations, with a clean experimental signature that allows determining precisely the mass of the (invisible) recoiling dark matter candidate \Suuddss.
It should be noted, however -- since the novel scheme here is based on multi-nucleon interactions -- that further theoretical work will be needed to assess the constraining power of any upper limits in the event that a signal is not observed. 

\nocite{*}

\bmhead{Acknowledgments}
This research of GRF has been supported by National Science Foundation Grant No.~PHY-2013199 and the Simons Foundation. 
This work of GK was supported by Warsaw University of Technology within the Excellence Initiative: Research University (ID-UB) programme and the IDUB-POB-FWEiTE-1 project grant as well as the Polish National Science Centre under agreement no. 2022/45/B/ST2/02029, and by the Polish Ministry of Education and Science under agreement no. 2022/WK/06.

\bibliography{sn-bibliography} 

\end{document}